# Probing phonon softening in ferroelectrics by the scanning probe microwave spectroscopy

Mykola Yelisieiev[1*], Petro Maksymovych[2], and Anna N. Morozovska [1,3†]

[1] Taras Shevchenko National University of Kyiv, Volodymyrska street 64, Kyiv, 01601, Ukraine

[2] The Center for Nanophase Materials Sciences, Oak Ridge National Laboratory, Oak Ridge, TN 37831

[3] Institute of Physics, National Academy of Sciences of Ukraine,

46, pr. Nauky, 03028 Kyiv, Ukraine

**Abstract**

Microwave measurements have recently been successfully applied to measure ferroelectric materials on the nanoscale, including detection of polarization switching and ferroelectric domain walls. Here we discuss the question whether scanning probe microscopy (SPM) operating at microwave frequency can identify the changes associated with the soft phonon dynamics in a ferroic. The analytical expressions for the electric potential, complex impedance and dielectric losses are derived and analyzed, since these physical quantities are linked to experimentally-measurable properties of the ferroic. As a ferroic we consider virtual or proper ferroelectric with an optic phonon mode that softens at a Curie point. We also consider a decay mechanism linked to the conductance of the ferroic, and thus manifesting itself as the dielectric loss in the material. Our key finding is that the influence of the soft phonon dispersion on the surface potential distribution, complex impedance and dielectric losses are evidently strong in the vicinity (~ 10 – 30 K) of the Curie temperature. Furthermore, we quantified how the spatial distribution and frequency spectra of the complex impedance and the dielectric losses react on the dynamics of the soft phonons near the Curie point. These results set the stage for characterization of polar phase transitions with nanoscale microwave measurements, providing a complementary approach to well established electromechanical measurements for fundamental understanding of ferroelectric properties as well as their applications in telecommunication and computing.

[*] a.k.a. − Nicholas E. Eliseev

[†] Corresponding author: anna.n.morozovska@gmail.com

# I. INTRODUCTION

Experimental and theoretical studies of the lattice dynamics in ferroics provide key fundamental insight into ferroelectric properties and their advanced applications. Since the lattice phase transition leads to the instability of specific phonon vibration modes, static displacements of atoms at the phase transition correspond to frozen displacements of the soft phonon modes [1, 2]. Basic experimental methods, which contain information about the soft phonon modes and spatial modulation of the order parameter in ferroics are dielectric measurements [3], inelastic neutron scattering [4, 5, 6], X-ray [7, 8, 9], Raman [10] and Brillouin [11] scatterings, and hypersound spectroscopic measurements [11].

Since no reliable microscopy techniques with frequencies up to THz (soft phonon range) and with high (nanoscale and below) resolution exist, it is important to study the frequency range by available probing methods. Note that THz microscopy has been well-studied, but it is a near-field scanning optical microscopy (NSOM) method, thus there is a light-in and light-out mechanisms, and the tip acts as an antenna [12, 13]. However, NSOM resolution has not yet reached the sub-micrometer range. A nanoscale resolution can be reached in different types of scanning probe atomic force microscopy, such as Kelvin probe and piezoresponse force microscopy (PFM), which are of particular importance for local probing the surface potential, elastic displacement [14, 15, 16] and polarization dynamics [17] of various ferroelectrics. However, PFM excitation frequency range is 10 KHz – 10 MHz, which is too low for the local probing of soft optical phonons in ferroelectrics in the THz range.

On the other hand, microwave scanning probe microscopy, where the tip is a leaky resonator [18], covers the range up to a few GHz (see topical reviews [19, 20, 21] and the monograph [22]). Nanoscale probing of the material electrodynamic properties at microwave frequencies is of great interest for materials science, condensed matter physics and device engineering [21]. Microwave microscopy can be in use for versatile applications [19], e.g., for study the electrophysical properties of materials, such as the electric conductivity, dielectric permittivity and dielectric losses at radio-frequency range. In addition to the nanoscale mapping of permittivity and conductivity, microwave microscopy is used for precise measurements of semiconducting and photosensitive materials, ferroelectric domains and domain walls, acoustic-wave systems and quantum materials with strong electron correlations [21, 23].

Important for this work, that the microwave probing provides nanoscale insights of the ferroelectric domain walls (FDWs) microwave conductance [24, 25], and the reversal process of a ferroelectric spontaneous polarization by an electric field. The applications of microwave microscopy for ferroelectrics can be found in modern nanoelectronics, in particular for probing of information processing devices [26], such as FeRAM memory cells and different logic units, where it is important to control the electric conductivity of the studied material. Moreover, the giant microwave conductivity of a "nominally silent"

FDWs [27] can open unexplored perspectives for their use in nanoelectronic devices based on the FDWs, such as memories, transistors, and rectifiers. The scanning microwave impedance microscopy (sMIM) that spatially resolves the ferroelectric DWs in a GHz range [26, 28], can open a new avenue to implement the FDWs in radio-frequency nanodevices.

Proper ferroelectrics remain tunable in the microwave range (which can be used for 5G networking [29, 30, 31]), but due to the collective response of polarizability, rather than to direct excitation of soft phonons. Following [19], there is a soft mode possible on the wall due to its vibration, although this phenomenon likewise constitutes collective response of polarization but in a lower-dimensional volume. Therefore, it is intriguing to consider mechanisms by which microwave measurements can probe phonon properties directly.

It is important to note that the image formation in the microwave scanning probe microscopy can be considered within a classical continuous medium approach, which offers possibilities for analytical calculations in various ferroics within e.g., decoupling approximation [32, 33, 34] well-elaborated for classical ferroelectrics [35]. In this work we focus on calculating the electric potential, whose properties define the response of the microwave scanning probe microscopy. We consider a ferroic (virtual or proper ferroelectric) with an optical phonon mode that is softened at Curie temperature. The problem statement is described in **Section II.** The analytical solution for the potential is derived and analyzed in **Section III**. In **Section IV** we studied the properties of the surface potential, namely in **subsection A** we considered the potential itself, in **subsection B** we evaluated and analyzed the complex impedance frequency spectra, and in **subsection C** we calculate and analyze the dielectric losses in the ferroic, and their link to the conductance of the material. Here we did not consider the losses associated with the possible motion of the FDWs in ferroelectrics near the Curie temperature $T_C$, since the motion is critically slowed down around $T_C$, and so the losses can be regarded negligibly small at GHz frequency. Since the domain structure is absent for improper ferroelectrics and paraelectrics, our results are even more valid for these materials. The surface potential, impedance and losses are linked to experimentally-measurable properties of a given ferroic. In **Section V** we summarize how the spatial distribution and frequency spectra of the calculated properties can help to determine the behavior of the soft phonon mode near the Curie point. Also, we have found certain spectrum peculiarities of the surface potential, impedance and losses near the Curie temperature, which can be useful for the analysis of the microwave spectroscopy data.

## II. PROBLEM STATEMENT

Let us study the question whether the scanning probe microscope (SPM), can register any changes of the electric potential and complex dielectric permittivity associated with electric charges redistribution at the ferroic surface. These phenomena are related with the oscillations of the tip effective charge at a microwave frequency (0.5 – 50) GHz and for different temperatures. There are many effective charge models [36, 37], as well as any axially-symmetric conducting tip can be modeled more rigorously by a set of image charges $Q_i$ at distances $d_i$ [38].

So, let us consider an oscillating effective charge $Q(t)$ suspended at the distance $d$ from the boundary between the media "1" and "2" (see **Fig. 1**). Of course, the charge redistribution has both vertical and lateral components, but hereafter we consider only the vertical one. The frequency of the charge oscillations (which varies from hundreds of MHz to tens of GHz,) is still much smaller than the optical frequencies. The distance $d$ is typically smaller than several nm.

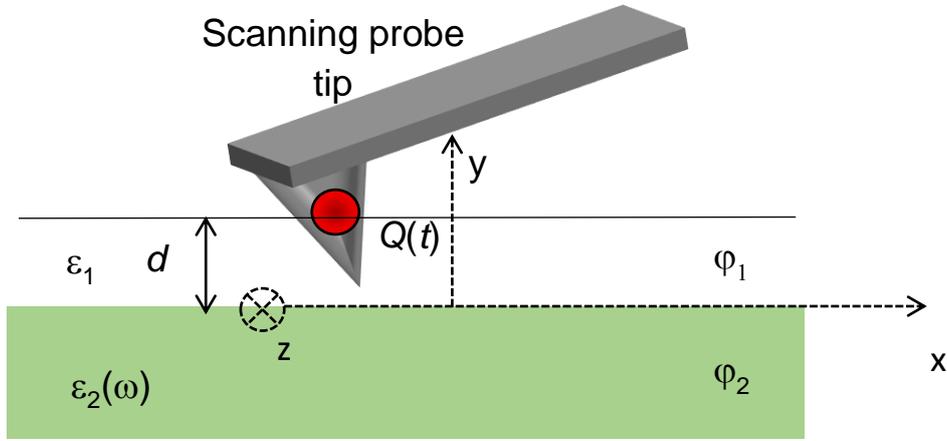

**Fig. 1.** An oscillating effective charge $Q(t)$ is suspended above an ambient-ferroic boundary at a distance $d$ from it.

The medium 1 is an isotropic dielectric (air or vacuum) without any dispersion, and the medium 2 is a ferroic, such as a proper or incipient ferroelectric, paraelectric, or a ferroelectric relaxor with an anisotropic dispersion of soft optic phonons. Their tensors of relative dielectric permittivity are shown below:

$$\varepsilon_{ij}^{(1)} \cong \begin{bmatrix} \varepsilon_e & 0 & 0 \\ 0 & \varepsilon_e & 0 \\ 0 & 0 & \varepsilon_e \end{bmatrix}, \quad \varepsilon_{ij}^{(2)} = \begin{bmatrix} \varepsilon_\perp & 0 & 0 \\ 0 & \varepsilon_\perp & 0 \\ 0 & 0 & \varepsilon_\parallel(\omega) \end{bmatrix}, \quad \varepsilon_\parallel(\omega) = \chi_\infty \frac{\omega_{lo}^2 - \omega^2 - 2i\gamma\omega}{\omega_{to}^2 - \omega^2 - 2i\gamma\omega}. \quad (1a)$$

The expression for $\varepsilon_\parallel(\omega)$ corresponds to the model for the Lorentzian oscillator with losses [39], where $\omega_{lo}^2$ corresponds to a longitudinal phonon mode; it frequency is of (0.2 – 2) THz order of magnitude for most paraelectric and ferroelectric materials [4, 2, 40, 41], and is almost temperature-independent. The asymptotic value $\chi_\infty \cong (3 - 9)$ corresponds to the high-frequency limit of dielectric permittivity, that, in fact is equal to the background permittivity [42]. The frequency $\omega_{to}$ is a transverse soft optical phonon mode, which is temperature dependent [43], as is shown below:

$$\omega_{to}^2 \cong \omega_f^2 \begin{cases} 1 - \frac{T}{T_C}, & \frac{T}{T_C} \leq 1, \\ \frac{T}{T_C} - 1, & \frac{T}{T_C} \geq 1. \end{cases} \quad (1b)$$

where $T_C$ is either virtual or real ferroelectric Curie temperature, $T$ is the absolute temperature. Since the dispersion of this mode is especially important in ferroics near the Curie temperature [44, 45, 46], below we consider this temperature range.

In order to relate the constants $\omega_{lo}^2$ and $\omega_f^2$ we use the Lyddane-Sachs-Teller (LST) relation, $\frac{\omega_{lo}^2}{\omega_{to}^2} = \frac{\varepsilon_{st}}{\chi_\infty}$ [47], where $\chi_\infty \cong (3-10)$ [48] and $\varepsilon_{st} \cong \varepsilon_f \left|1 - \frac{T}{T_C}\right|^{-1}$ is a static relative dielectric permittivity of a ferroelectric (or paraelectric) that diverges at $T \to T_C$. We obtain that

$$\omega_f = \omega_{lo}\sqrt{\frac{\chi_\infty}{\varepsilon_f}}. \quad (1c)$$

Since $30 \leq \varepsilon_f \leq 3000$ for most incipient (like SrTiO3) and proper (like BaTiO3, PbTiO3) ferroelectrics, and ferroelectric relaxors systems (like PMN-PT), the inequality $0.03 \leq \frac{\omega_f}{\omega_{lo}} \leq 0.5$ is valid.

Meanwhile, $\gamma$ is a decay factor related with several mechanisms, such as Khalatnikov relaxation of polarization (acoustic phonons) and different pinning effects [49]. The order of magnitude of this value can be rather small or high, which is determined by the relaxation times. The Khalatnikov relaxation time typically varies in the range $(10^{-9} - 10^{-6})$ seconds for temperatures far from $T_C$. Below we will regard that $\gamma$ values can vary in the range $(10^9 - 10^6)$Hz.

The primary goal of our work is to find the potential distribution near the surface of the studied media. First, we need to look which equations the displacement field satisfies, and determined the possibility to transfer to the Poisson equation for the potential. The electric displacement satisfies the following equation system (2), which is the Maxwell equation in a differential form combined with the response integral [50]:

$$\begin{cases} div\boldsymbol{D} = Q(t)\,\delta(x)\delta(y)\delta(z-d), \\ \boldsymbol{D}(x,y,z,t) = \boldsymbol{P}_S + \varepsilon_0 \int_{-\infty}^{0} \varepsilon(t-\tau)\boldsymbol{E}(x,y,z,\tau)d\tau, \end{cases} \quad (2)$$

where the measurement protocol is $Q(t) = \int_0^\infty q(\omega)e^{i\omega t}d\omega$, and its spectral density $q(\omega)$ corresponds to a specific protocol of the voltage applied to the SPM tip. We suppose that $q(\omega)$ is nonzero in the frequency range $\omega < \omega_{max}$, where the maximal frequency $\omega_{max}$ is much smaller than the optical frequency $\omega_{opt}$. The spontaneous polarization $\boldsymbol{P}_S$ exists in proper ferroelectrics below Curie temperature. Its value is proportional to $\sqrt{T - T_c}$ for the proper ferroelectrics with the second-order phase transition.

Hereinafter we consider either improper ferroelectrics and quantum paraelectrics, where $\boldsymbol{P}_S \equiv 0$, or the second-order proper ferroelectrics near the Curie temperature $T_C$. At the temperatures above $T_C$ $\boldsymbol{P}_S$ is absent. At the temperatures slightly below $T_C$ the dynamics of $\boldsymbol{P}_S$ becomes "sluggish" due to the critical slowing down effect [51]. The effect is related with the fact that the polarization relaxation time is inversely proportional to $|T - T_C|$ in the framework of Landau-Khalatnikov model [49, 52]. In other words, we regard that the dynamics of spontaneous polarization $\boldsymbol{P}_S(\boldsymbol{r})$ related with e.g., the FDWs motion, has characteristic relaxation times much smaller than 1 - 100 microseconds, that is valid for proper ferroelectrics due to the omnipresent lattice pinning and intrinsic defects [17]. Hence, we assume that the "slow" dynamics of $\boldsymbol{P}_S(\boldsymbol{r})$ is insensitive to the "fast" GHz oscillations the tip voltage. The assumption becomes invalid far from the Curie temperature, when the "trembling" of the FDWs can be observed by a microwave SPM operating at GHz frequencies [27].

The boundary conditions for the problem (2) are:

$$D_z^{(1)} - D_z^{(2)}\Big|_{z=0} = \sigma_S, \quad E_x^{(1)} = E_x^{(2)}\Big|_{z=0}, \quad E_y^{(1)} = E_y^{(2)}\Big|_{z=0} \quad (3)$$

Here $\sigma_S$ is the density of a sluggish surface free charge, which screens a polarization bound charge in order to prevent the giant depolarization field caused by the bound charge [53, 54]. The nature of the screening charges is external electrons, protons or hydroxyl groups abundant in the medium 1 at normal ambient conditions and humidity 30% or more, or the internal band bending appearing in the ferroic medium 2 under the ultra-high vacuum in the medium 1 [55]. As a rule, the condition of complete screening, $\sigma_S = -P_S$, is valid with a high accuracy at electrically-open ferroelectric surfaces without special passive layers [17]. Below we regard that the condition $\sigma_S = -P_S$ is hold due to the presence of ambient free charges $\sigma_S(\boldsymbol{r})$, which completely screen the bound charge $\boldsymbol{P}_S(\boldsymbol{r})$ outside the ferroelectric.

### III. APPROXIMATE ANALYTICAL SOLUTIONS

We can neglect the influence of magnetic fields and put $rot\vec{E} = 0$, because $\omega_{max} \ll \omega_{opt}$, and introduce a scalar electrostatic potential $\varphi$. After a Fourier transform on time $t$ we obtain the electrostatic problem with boundary conditions:

$$\begin{cases} \varepsilon_e \Delta \varphi^{(1)}(\omega, x, y, z) = -\frac{\pi}{\varepsilon_0} q(\omega)\, \delta(x)\delta(y)\delta(z-d), & z > 0, \\ \varepsilon_\perp \Delta_\perp \varphi^{(2)}(\omega, x, y, z) + \varepsilon_\parallel(\omega)\frac{\partial^2}{\partial z^2}\varphi^{(2)}(\omega, x, y, z) = 0, & z < 0, \\ \varphi^{(1)}\big|_{z=0} = \varphi^{(2)}\big|_{z=0}, \\ \varepsilon_e \frac{\partial \varphi^{(1)}}{\partial z}\bigg|_{z=0} = \varepsilon_\parallel(\omega)\frac{\partial \varphi^{(2)}}{\partial z}\bigg|_{z=0}. \end{cases} \quad (4)$$

The solution of Eq.(4) was found by the means of a Fourier transform from **r**-space to **k**-space [56]. For the upper half-space (Fig. 1), $z > 0$, the potential is:

$$\varphi^{(1)}(t, r, z) = \frac{1}{\varepsilon_0 \varepsilon_e} \mathrm{Re}\left[\int_0^{\omega_{max}} \left(\frac{1}{\sqrt{(z-d)^2+r^2}} - \frac{\sqrt{\varepsilon_\perp \varepsilon_\parallel(\omega)} - \varepsilon_e}{\sqrt{\varepsilon_\perp \varepsilon_\parallel(\omega)} + \varepsilon_e}\frac{1}{\sqrt{(z+d)^2+r^2}}\right) q(\omega) e^{i\omega t} d\omega\right]. \quad (5a)$$

Hereinafter a polar radius $r^2 = x^2 + y^2$ is introduced due to the axial symmetry of the system, and the dispersion law for $\varepsilon_\parallel(\omega)$ is given by Eqs. (1). The potential in the lower half-space ($z < 0$) of the ferroic volume is equal to

$$\varphi^{(2)}(t,r,z) = \frac{1}{\varepsilon_0} \text{Re}\left[\int_0^{\omega_{max}} \left(\frac{2}{\varepsilon_e + \sqrt{\varepsilon_\perp \varepsilon_\parallel(\omega)}} \frac{1}{\sqrt{r^2 + \left(d - z\sqrt{\varepsilon_\perp/\varepsilon_\parallel(\omega)}\right)^2}}\right) q(\omega) e^{i\omega t} d\omega\right]. \quad (5b)$$

Note that the convergency condition of integrals (5) is $\text{Re}\left[\sqrt{\frac{\varepsilon_\perp}{\varepsilon_\parallel(\omega)}}\right] > 0$. The surface potential $\varphi^{(1)}(t,r,0) = \varphi^{(2)}(t,r,0)$, since the electric potential is continuous at the ferroic-ambient interface. Below we will use the designation $\varphi(t,r,0)$ for the surface potential.

We also note, that the first term in Eq.(5a), proportional to $\frac{1}{\sqrt{(z-d)^2+r^2}}$, is related to presence of charge, while the second term, proportional to $\frac{1}{\sqrt{(z+d)^2+r^2}}$, corresponds to its dynamic image. The only term in Eq.(5b), proportional to $\frac{1}{\sqrt{r^2+\left(d-z\sqrt{\varepsilon_\perp/\varepsilon_\parallel(\omega)}\right)^2}}$, is related with the image charge dynamics. According to Eqs. (5), to probe a pronounced dispersion by the SPM tip, we need a ferroic material with a small $\chi_\infty$, high decay $\gamma$ and significant difference between the longitudinal and transverse optical phonon frequencies, $\omega_{lo}^2 - \omega_{to}^2$.

The following dimensionless parameters are introduced hereafter:

$$\widetilde{\omega}_m = \frac{\omega_{max}}{\omega_{lo}}, \quad \widetilde{\omega} = \frac{\omega}{\omega_{lo}}, \quad \widetilde{\gamma} = \frac{\gamma}{\omega_{lo}}, \quad \widetilde{\omega}_{to} = \frac{\omega_{to}}{\omega_{lo}} = \frac{\omega_f}{\omega_{lo}}\sqrt{\theta}, \quad (6a)$$

$$\theta = \left|\frac{T-T_C}{T_C}\right|, \quad x' = \frac{x}{d}, \quad y' = \frac{y}{d}, \quad r' = \frac{r}{d}, \quad z' = \frac{z}{d}. \quad (6b)$$

Note that the deviation $\theta$, proportional to the absolute value $|T - T_C|$, is always positive, and so below we consider the temperatures $T$ higher and lower the Curie temperature $T_C$.

Since the soft phonon frequency $\omega_{lo} \sim (0.5 - 2)10^{12}$Hz, the microwave frequency $\omega_{max} \sim (0.5 - 50)10^9$Hz and the decay factor $\gamma \sim (10^9 - 10^6)$Hz, the dimensionless parameters $\widetilde{\gamma}$ and $\widetilde{\omega}_m$ vary in the range, $10^{-6} \leq \widetilde{\gamma} \leq 10^{-3}$ and $10^{-3} \leq \widetilde{\omega}_m \leq 10^{-1}$. Also, we consider the ratio and $0.03 \leq \frac{\omega_f}{\omega_{lo}} \leq 0.5$ for the reasons discussed above, and regard that the relative deviation from real or virtual Curie temperature $T_C$ is within 10%, corresponding to the range (10 – 60) K. So, the relative deviation $\theta$ from real or virtual Curie temperature is the range $0 \leq \theta \leq 0.1$. One can consider a wider range of temperature deviations, and have found that the most peculiar changes in the curves shape are concentrated around the 10%. But results presented below are valid only for small $\theta$, for which we can expect that the response of the domain structure is still negligible due to the proximity to the Curie temperature.

## IV. RESULTS AND DISCUSSION
### A. Surface potential properties

We have studied the dependence of the surface potential Fourier image $\varphi^S(\omega, r, 0)$ on the frequency $\omega$ of the charge modulation. We have considered the measurement protocol $Q(t)$ a harmonic function, and the corresponding transfer (or measurement) function to be equal to $q(\omega) = q_0 \delta(\omega - \omega_0)$, where $\omega_0$ is a constant value in GHz range. In particular, below we analyze the spatial distribution and the temperature dependence of the potential.

As one can see from **Fig. 2**, the spectrum of the surface potential strongly depends on the temperature in the vicinity of Curie temperature, $\theta = 0$. Indeed, as is shown on the **Fig. 2(a)**, there occurs an antiresonance in the form of a seagull-shape minimum on the frequency dependence. The minimum is caused by the soft optical phonon dynamics and its frequency position depends on the temperature deviation $\theta$ as $\widetilde{\omega} = \widetilde{\omega}_{to}(\theta) = \frac{\omega_f}{\omega_{lo}}\sqrt{\theta}$. This minimum corresponds to the peculiarity on the frequency dependence of the real part of dielectric permittivity, which is shown on the inset. When the temperature approaches the Curie point, the antiresonance minimum shifts towards lower frequencies proportionally to $\sqrt{\theta}$ and disappears completely at the Curie point $\theta = 0$. This effect can be used in various experiments, for example in order to measure the soft phonon mode frequencies and determine the Curie point of a given material.

Note that **Fig. 2(a)** is plotted for a negligibly small decay value, $\widetilde{\gamma} = 10^{-6}$. The natural tendency is that the minima of the surface potential become much more diffuse for a higher decay, and the potential is significantly smaller. From **Fig. 2(b)**, plotted for a much higher decay value, $\widetilde{\gamma} = 10^{-3}$, the minima flatten out, but are still pronounced and correspond well to the sharp maxima of the dielectric permittivity imaginary part. Because $\widetilde{\gamma} \ll 1$, the minima position still correspond to the frequency value $\widetilde{\omega} = \frac{\omega_f}{\omega_{lo}}\sqrt{\theta}$ and shift towards lower frequencies with $\theta$ decrease. Actually, the surface potential minimum moves within the dimensionless frequency range $0.015 \leq \widetilde{\omega} \leq 0.035$ under the temperature variation in the range $0.01 \leq \theta \leq 0.05$. Note that corresponding $\omega_0$ range (10 – 40) GHz is now in the upper range of microwave measurements.

The spatial distribution of the surface potential has a pronounced maximum in the position under the tip apex ($r = 0$), and it decays rapidly with the distance from this position [see **Fig.2(c)**]. Note that the potential changes significantly even with small temperature changes in the range close to the Curie point. Indeed, with the temperature increase the maximum flattens out and is virtually zero at the Curie temperature. The decay coefficient doesn't influence the spatial distribution significantly, if it is changed in physically reasonable limits $10^{-6} \leq \widetilde{\gamma} \leq 10^{-3}$. Within the actual range of $\widetilde{\gamma}$ the temperature dependance of the surface potential is monotonic, it decreases with $\theta$ decrease reaching minimum in the Curie point, $\theta = 0$, and also decreases with the distance $x'$ [see **Fig.2(d)**].

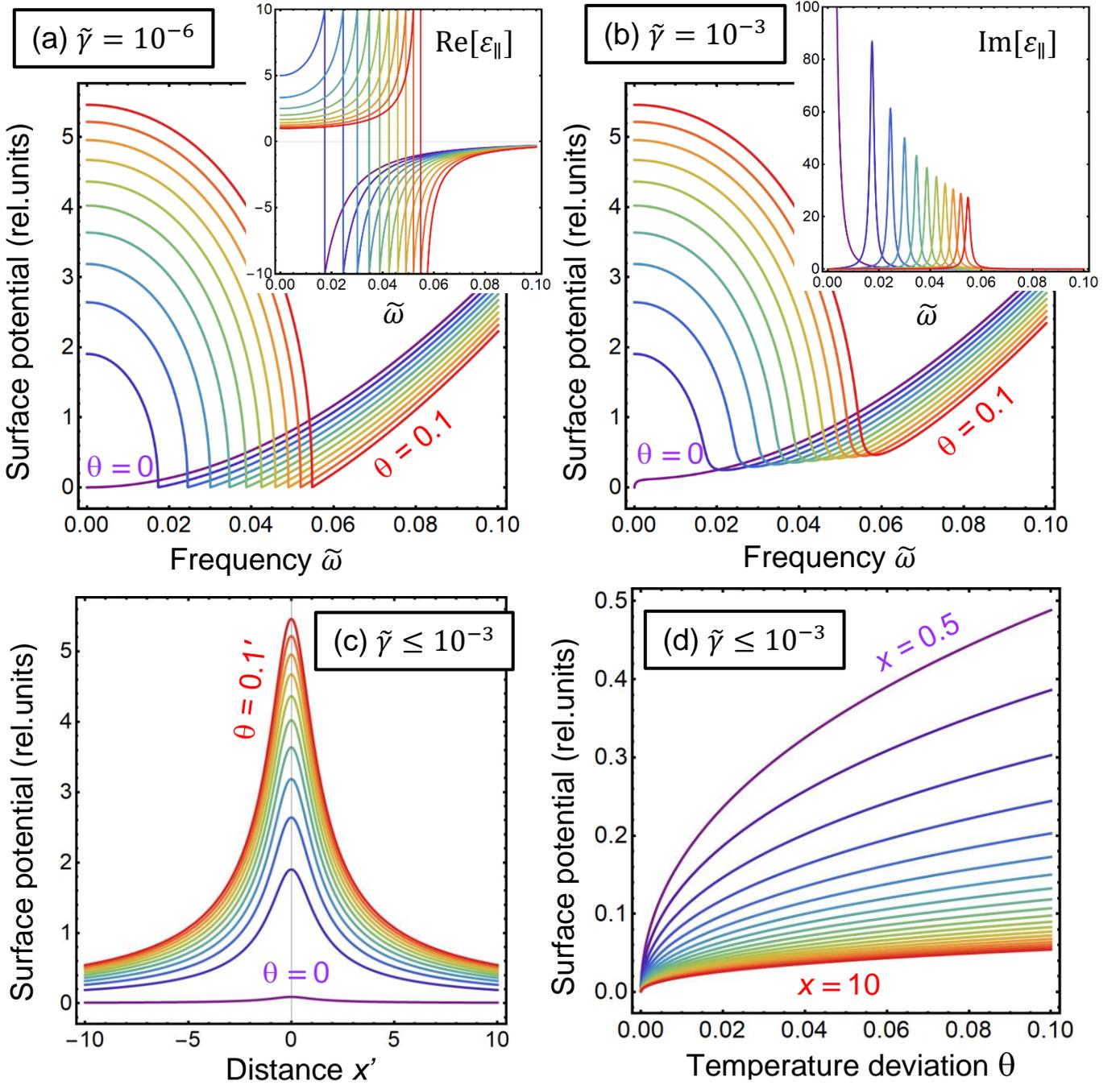

**Figure 2.** The frequency spectrum of the surface potential $\varphi(\widetilde{\omega})$ under the tip apex ($r' = 0$) calculated for dimensionless decay constants $\widetilde{\gamma} = 10^{-6}$ (a) and $\widetilde{\gamma} = 10^{-3}$ (b), deviation $\theta$ from Curie temperature ranging from $\theta = 0.1$ (red curves) to $\theta = 0$ (violet curves) with a step of 0.01. (c) The surface potential distribution vs. the distance $x'$ from the tip apex calculated for the harmonic excitation with frequency $\widetilde{\omega}_0 = 0.001$, $\widetilde{\gamma} \leq 10^{-3}$ and $\theta$ ranging from 0.1 (red curve) to 0 (violet curve) with a step of 0.01. (d) The surface potential vs. the $\theta$ calculated for the harmonic excitation with frequency $\tilde{\omega}_0 = 0.001$, $\tilde{\gamma} \leq 10^{-3}$ and distance $x'$ ranging from 0.5 (violet curve) to 10 (red curve) with a step of 0.5. Insets in the plots (a) and (b) show real and imaginary parts of the relative dielectric permittivity $\varepsilon_\parallel(\omega)$ calculated for $\tilde{\gamma} = 10^{-6}$ and $\tilde{\gamma} = 10^{-3}$, respectively. The ratio $\frac{\omega_f}{\omega_{lo}} = 0.17$ and $\chi_\infty = 3$.

### B. Capacitance of the SPM junction

To study the fingerprints of the soft phonon dynamics in the behavior of the nearly ideally insulating ferroic near Curie temperature by the microwave microscopy, one needs to calculate the effective capacitance of the system. We consider the case when the voltage $U$ is applied between the tip and the substrate; and the tip apex is almost touching the surface of a thick ferroic film. As we modeled the tip by an effective charge $Q$ at distance $d$ from the ferroic surface, we need to relate the parameters of the charge with the tip geometry and the applied voltage. The proportionality coefficient gives us the effective capacitance. There are several models of an effective charge, described in details in Ref.[37] and summarized in Table I, therein. The most well-known are sphere-plane, disk-plane, effective point charge and capacitance models. The choice of the specific model depends on the spatial region, where we want to model the probe field with the highest accuracy. Since we are interested in the local microwave probing with minimal influence from the stray fields [19], the most suitable is the disk-plane model of the tip, which reproduces the electric field of the conductive flattened apex, represented by the disk of radius $R_0$ that touches the surface. The total charge equals to $Q = C_t U$. For a thick film with a thickness of $h \gg R_0$ the capacitance $C_t$ and effective distance $d$ are [37]:

$$C_t \approx 4\varepsilon_0\left(\varepsilon_e + \sqrt{\varepsilon_\perp \varepsilon_\parallel(\omega)}\right)R_0, \qquad d \approx \frac{2}{\pi}R_0. \qquad (7)$$

Within the disk-plane model framework the field structure is adequately described in the most part of the piezoresponse volume, since $\phi(r \leq R_0, 0) \approx U$ and $\phi(r \gg R_0) \sim Q/r$. However, in the considered case $\varepsilon_\parallel(\omega)$ is a complex value, and so the effective capacitance is also complex. Note that the "dielectric" factor, $\varepsilon_e + \sqrt{\varepsilon_\perp \varepsilon_\parallel(\omega)}$, is present the surface potential given by expression $\varphi(t, r, 0) = \frac{1}{\varepsilon_0} \text{Re}\left[\int_0^{\omega_{max}} \left(\frac{2}{\varepsilon_e + \sqrt{\varepsilon_\perp \varepsilon_\parallel(\omega)}} \frac{1}{\sqrt{r^2 + d^2}}\right) q(\omega) e^{i\omega t} d\omega\right]$ [see the comment to Eqs.(5)].

Below we analyze the spectral properties of the complex impedance $Z(\omega) = \frac{1}{\omega C_t}$, namely its real and imaginary parts:

$$\text{Re}[Z(\omega)] = \frac{1}{4\varepsilon_0 R_0 \omega} \text{Re}\left[\frac{1}{\varepsilon_e + \sqrt{\varepsilon_\perp \varepsilon_\parallel(\omega)}}\right], \qquad \text{Im}[Z(\omega)] = \frac{1}{4\varepsilon_0 R_0 \omega} \text{Im}\left[\frac{1}{\varepsilon_e + \sqrt{\varepsilon_\perp \varepsilon_\parallel(\omega)}}\right]. \qquad (8)$$

It is evident that the complex impedance is proportional to the surface potential (5), measured under the tip in the point $r' = 0$.

The frequency dependence of the real and imaginary parts of the complex impedance $Z(\omega)$ are shown in **Fig. 3** for different temperature values. The real part, $\text{Re}[Z(\omega)]$, has a pronounced seagull-shape minimum, corresponding to the soft phonon frequency, as shown in **Fig. 3(a)** for a relatively small decay coefficient $\tilde{\gamma}$. This minimum moves to lower frequencies and disappears completely when the temperature tends to the Curie point, $\theta = 0$.

The minimum shape becomes smoother for high decay constants [see **Fig. 3(b)**]. Since $\tilde{\gamma} \ll 1$ in the actual range of material parameters, the frequency position of $\text{Re}[Z(\omega)]$ minimum is the same as for the surface potential shown in **Fig. 2(a)-(b)**. The minimum is related with soft optical phonons; it corresponds to the frequency value $\tilde{\omega} = \frac{\omega_f}{\omega_{lo}}\sqrt{\theta}$ and shifts towards lower frequencies with $\theta$ decrease in the range $0 \leq \theta \leq 0.05$. It seems very important for experimental observation of the $\text{Re}[Z(\omega)]$ minima, that its frequency position $0.015 \leq \tilde{\omega} \leq 0.035$ corresponds to the (10 – 40) GHz range potentially "reachable" in microwave spectroscopy experiments.

The imaginary part, $\text{Im}[Z(\omega)]$, shown in **Fig. 3(b)** for small decay coefficient $\tilde{\gamma}$, has a sharp falling step at the position of $\text{Re}[Z(\omega)]$ minimum, $\tilde{\omega} = \frac{\omega_f}{\omega_{lo}}\sqrt{\theta}$. The step shifts towards lower frequencies with $|\theta|$ decrease. After falling down from the step, $\text{Im}[Z(\omega)]$ monotonically decreases with increasing of the external field frequency and then saturates at high frequencies. With the increase of the decay constant the step diffuses and the transition between the saturation mode and the decrease becomes much smoother [see **Fig. 3(d)**], and in both cases the temperature increase shortens the saturation part of $\text{Im}[Z(\omega)]$. It is notable that the step disappears completely at the Curie point [see the violet curve in **Fig. 3(d)**].

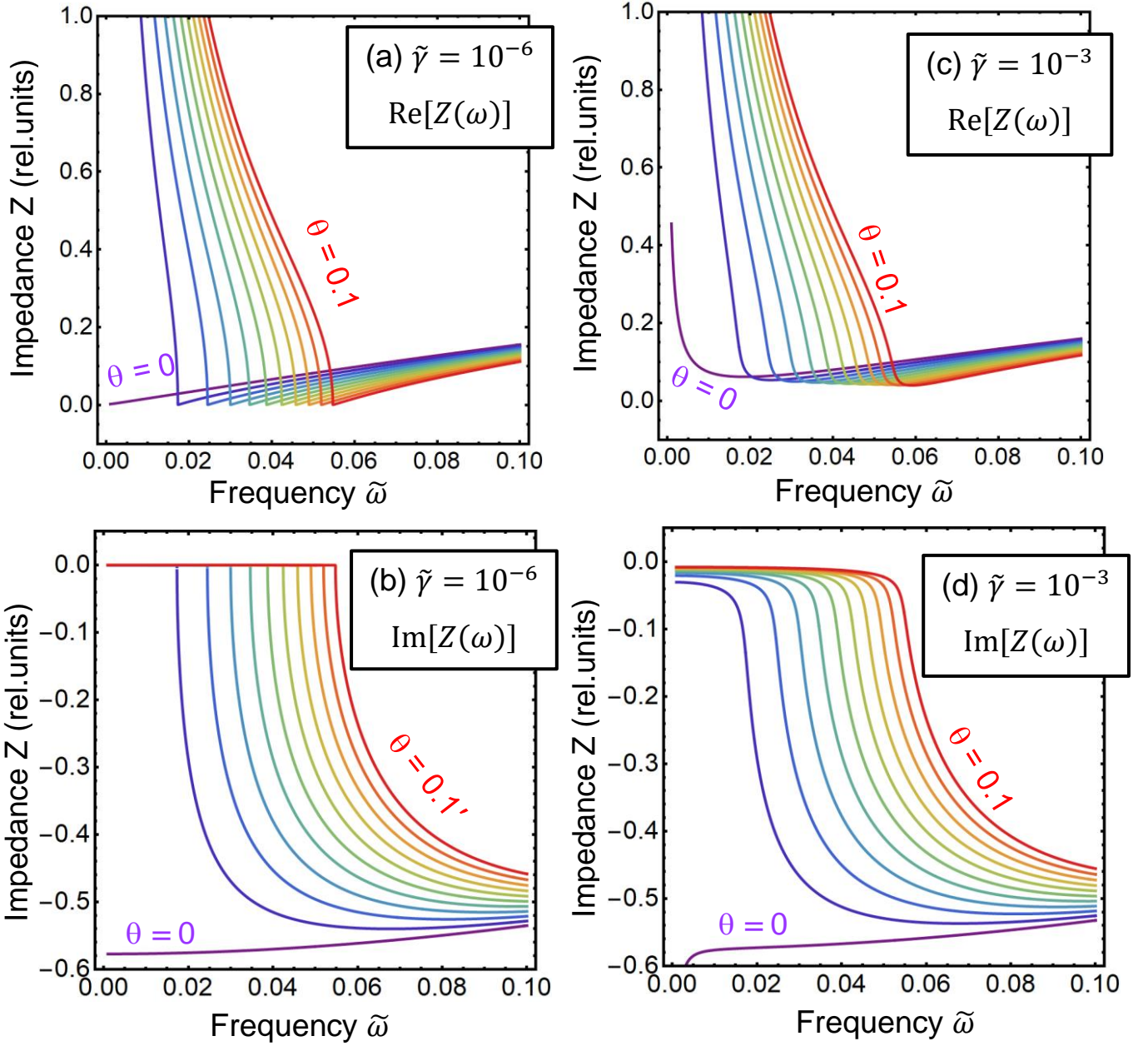

**Figure 3.** The dependence of the complex impedance $Z(\omega)$ real **(a,c)** and imaginary **(b,d)** parts on the frequency $\tilde{\omega}$ calculated for decay constants $\tilde{\gamma} = 10^{-6}$ **(a, b)** and $\tilde{\gamma} = 10^{-3}$ **(c,d)** and ten dimensionless deviations $\theta$ from the Curie temperature ranging from $\theta = 0.1$ (red curves) to $\theta = 0$ (violet curves) with a step of 0.01. The ratio $\frac{\omega_f}{\omega_{lo}} = 0.17$ and $\chi_\infty = 3$.

### C. Dielectric losses

Let us estimate the dielectric losses in the media, because they are important for the experimental studies as it is shown in Ref.[19]. This physical value quantifies the dissipation of the electromagnetic (EM) energy due to dipolar relaxation or the excitation of phonon modes. The practically important part is that the dielectric loss, defined by the imaginary part of the dielectric permittivity, is related with the conductivity of a material. The conductivity can be measured via an equivalent circuit. The general formula for the power of the said losses, as stated in the textbook [57], is shown below:

$$\frac{d\mathcal{E}}{dt} = -\varepsilon_0 \iiint_V \vec{E} \frac{\partial \vec{D}}{\partial t} dV. \qquad (9)$$

As one can see, the formula (9) is an expression for the time derivative of the EM energy in the media. The total energy per unit length can be written as follows, in the form of an integration on time:

$$\mathcal{E} = \Delta x \int_{-\infty}^{+\infty} dt \iint_{-\infty}^{+\infty} \rho_Q dy dz. \qquad (10)$$

Here $\rho_Q = -\varepsilon_0 \vec{E} \frac{\partial \vec{D}}{\partial t}$ is the integrand in Eq. (9). As it is derived in **Suppl. Mat.** (see Eqs.(S.40)-(S.42) in Ref.[56]), expression (10) can be rewritten as follows:

$$\mathcal{E} = \iiint_V \rho_E dV, \quad \rho_E = -\frac{1}{2\pi} \int_0^{+\infty} \omega |\vec{\nabla}\psi^{(2)}(\omega,r,z)|^2 \operatorname{Im}[\varepsilon_\parallel(\omega)] d\omega, \qquad (11)$$

Where the Fourier-image of the potential has the following form

$$\psi^{(2)}(\omega,r,z) = \left( \frac{2}{\varepsilon_e + \sqrt{\varepsilon_\perp \varepsilon_\parallel(\omega)}} \frac{1}{\sqrt{r^2 + (d-z\sqrt{\varepsilon_\perp/\varepsilon_\parallel(\omega)})^2}} \right) q(\omega) \qquad (12)$$

Using this expression, the expression for $\rho_E(\omega,r,z)$ can be transformed to the expression:

$$\rho_E(\omega,r,z) = -4\varepsilon_0 \frac{\omega \operatorname{Im}[\varepsilon_\parallel(\omega)] |q(\omega)|^2}{|\varepsilon_e + \sqrt{\varepsilon_\perp \varepsilon_\parallel(\omega)}|^2} \frac{r^2 + |d-z\sqrt{\varepsilon_\perp/\varepsilon_\parallel(\omega)}|^2 |\varepsilon_\perp/\varepsilon_\parallel(\omega)|}{\left|r^2 + (d-z\sqrt{\varepsilon_\perp/\varepsilon_\parallel(\omega)})^2\right|^3}. \qquad (13)$$

As one can see, the expression (13) gives the spatial distribution of the spectral density of the dielectric losses, which is proportional to $\operatorname{Im}[\varepsilon_\parallel(\omega)] \sim \gamma$. The surface distribution obtained from this formula ($z=0$) is important for the microscopy mapping of the conductivity.

The spectrum of the dielectric losses, shown in **Fig. 4(a)**, has a very sharp resonance peak, which corresponds to the soft phonon frequency of the ferroic. The peak moves to lower frequencies with the temperature deviation $\theta$, and the absolute value of the losses increases with $\theta$ increase. At the Curie point the resonance disappears completely, and is replaced with a monotonous decrease of the losses value, as can be seen from the inset in **Fig. 4(a)**.

With the increase of the decay constant $\gamma$, the peaks become much lower and wider, and their absolute value decreases with the temperature [compare **Fig. 4(c)** with **4(a)** allowing for the three-order difference in the scaling factors]. Because the physically reasonable range of decay constant is $\tilde{\gamma} \ll 1$, the frequency position of the $\rho_E(\omega,r,z)$ peak is the same as for the surface potential and impedance peculiarities shown in **Figs. 2** and **3**, respectively. As it should be, the peak of $\rho_E$ of is related with soft optical phonons; it corresponds to the frequency value $\tilde{\omega} = \frac{\omega_f}{\omega_{lo}} \sqrt{\theta}$ and shifts towards lower frequencies with $\theta$ decrease. It seems very important for experimental observation of the maximal dielectric losses that the peak frequency position $0.015 \leq \tilde{\omega} \leq 0.035$ corresponds to the (10 – 40) GHz range potentially "reachable" in microwave spectroscopy experiments.

The spatial distribution, which can be mapped on the experiment, shows a well-defined minimum located at a certain distance from the tip of the probe [see **Fig.4(c)**]. The absolute value of the losses at this maximum increase with the temperature [see **Fig.4(d)**]. The inset [see **Fig.4(c)**] shows that the mapping shows a defined "ring" of maximum loss values around the tip.

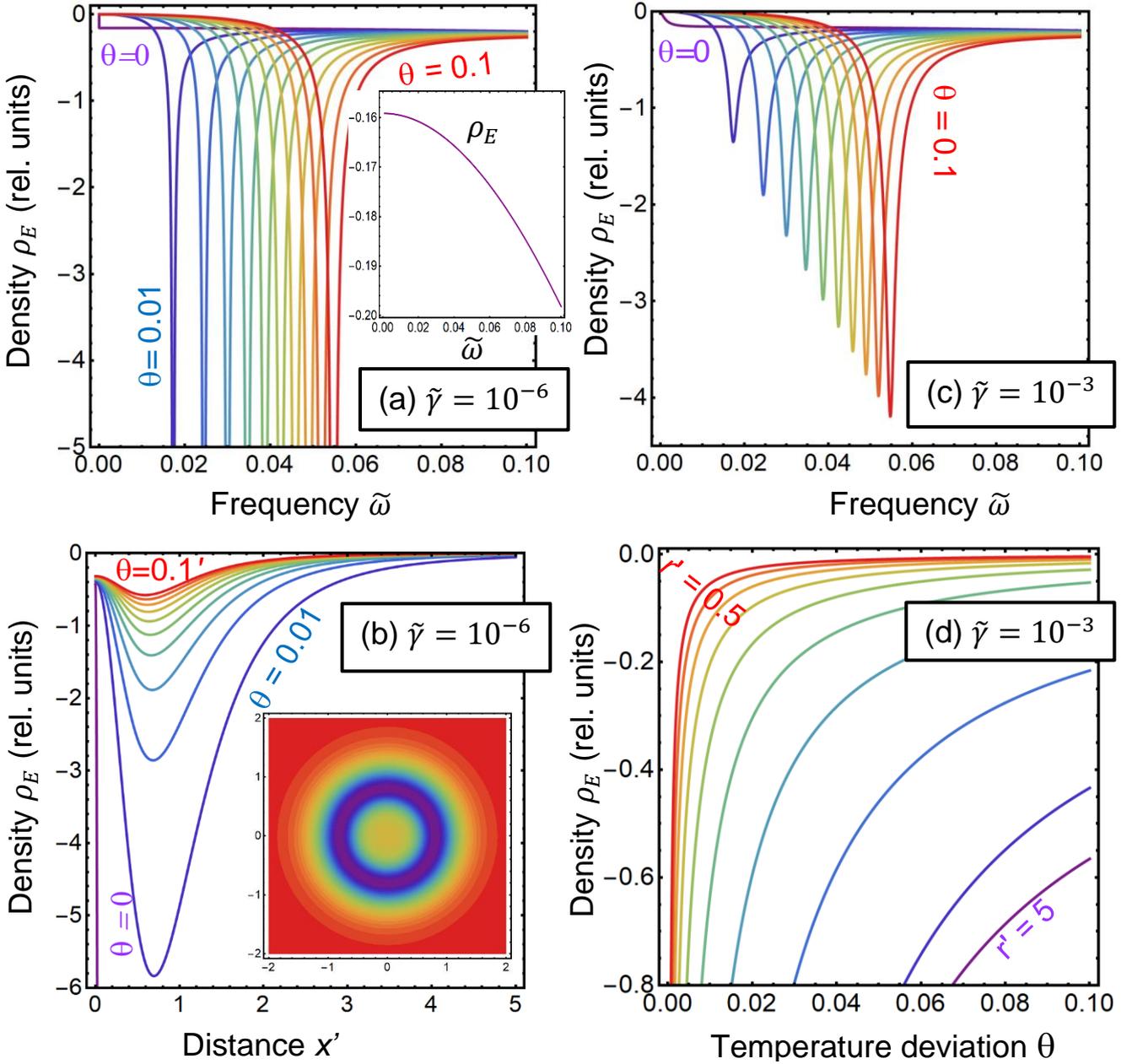

**Figure 4.** The dependence of the dielectric losses spectral density $\rho_E(\tilde{\omega}, r', z)$ on the dimensionless frequency $\tilde{\omega}$ (**a,c**), on the distance (**b**), and the temperature deviation $\theta$ (**d**) calculated for decay constants $\tilde{\gamma} = 10^{-6}$ (**a, b**) and $\tilde{\gamma} = 10^{-3}$ (**c,d**), and ten $\theta$ values ranging from 0.1 (red curves) to 0 (violet curves) with a step of 0.01 (**a, b, c**), and different distance $r'$ ranging from 0.5 to 5 **(d)**. All the calculations were done for the surface distribution, $z = 0$. The dimensionless radius $r'$ is equal to unity in the plots **(a)** and **(c)**. The modulation frequency is $\widetilde{\omega}_0 = 0.001$ for panels **(b)** and **(d)**. The inset in the plot **(a)** shows the spectrum $\rho_E(\widetilde{\omega})$ at the Curie temperature $\theta = 0$. The inset in the plot **(b)** shows the spectrum $\rho_E(\widetilde{\omega})$ in $\{x, y\}$-section near the Curie temperature, $\theta = 0.05$. Note that the scaling factors are $10^8$, $10^5$, $10^{12}$ and $10^9$ for the plots **(a), (c), (b)** and **(d)**, respectively. The ratio $\frac{\omega_f}{\omega_{lo}} = 0.17$ and $\chi_\infty = 3$.

## IV. CONCLUSION

The analytical solution for the electric potential, complex impedance and dielectric losses of a given ferroic (virtual or proper ferroelectric) is derived and analyzed, since these physical quantities are experimentally-measurable by a SPM operating at a GHz frequency. We consider the ferroic to be an incipient or a proper ferroelectric with an optic phonon mode that is softened at the Curie point and included a decay mechanism is linked to its conductance. Our analysis proves that the influence of the soft phonon dispersion on the surface potential distribution, complex impedance and dielectric losses frequency spectra is especially strong in the 10% vicinity of the Curie temperature. The result can be useful for the analysis of the microwave spectroscopy data. Specifically:

(a) A pronounced phonon dispersion can be probed by the SPM tip in a ferroic material with a small high-frequency dielectric permittivity (~3 or less), high decay constant and significant difference between the longitudinal and transverse optical phonon frequencies.

(b) Because the reasonable values of the decay constant are relatively small, the frequency position of the surface potential, complex impedance and dielectric losses features, such as seagull minima, steps and peaks, is the same as for all these features. All the features are related with soft optical phonons dynamics and shift towards lower frequencies when the temperature approaches Curie point.

(c) Since the frequency position of the dielectric losses and complex impedance features correspond to the (10 – 40) GHz range they are potentially "observable" in microwave spectroscopy experiments, and so our theoretical predictions can be verified experimentally.

(d) However, here we did not consider the losses associated with the possible motion of FDWs in a proper ferroelectric, as the motion was regarded too sluggish near the Curie temperature in comparison with fast oscillations the tip voltage at GHz frequency. Obtained results are even more valid for improper ferroelectrics and paraelectrics, where the domain walls are absent per se. For these cases the losses spectral density is proportional to the decay factor $\gamma$ that is linked to the material conductance, $\gamma \sim \text{Im}[\varepsilon_\parallel(\omega)]$. The conclusion is invalid far below the Curie temperature, when the response of the FDWs can be observed by a MIM operating at GHz frequencies [27].

**Acknowledgements.** Authors express their deepest gratitude to Prof. Dr. V.A. Kochelap (NASU), and Dr. Sergei V. Kalinin (ORNL) for stimulating discussions and fruitful ideas. A.N.M. work is supported by the National Academy of Sciences of Ukraine (grant ВЦ-1230, "New materials for nanoelectronics, ferrotronics and spintronics, and their research"). Analytical calculations are performed and visualized in the Mathematica 12.2 software (https://www.wolfram.com/mathematica). This material is based upon work (P.M.) supported by the Division of Materials Science and Engineering, Office of Science, Office of Basic Energy Sciences, U.S. Department of Energy, and performed in the Center for Nanophase Materials Sciences, supported by the Division of Scientific User Facilities. A portion of calculations was conducted at the Center for Nanophase Materials Sciences, which is a DOE Office of Science User Facility (CNMS Proposal ID: CNMS2021-B-00843).

**Authors' contribution.** A.N.M. and P.M. generated the research idea, and A.N.M. state the problem mathematically. M.Y. solved the problem analytically, generated figures and wrote corresponding original parts of the manuscript. A.N.M. and P.M. contribute to the introduction, results discussion, conclusions and manuscript improvement.

# Supplementary Materials to the "Probing phonon softening in ferroelectrics by the scanning probe microwave spectroscopy"

The electric displacement satisfies the system of equations (S.1), which is the Maxwell equation in a differential form combined with the response integral:

$$\begin{cases} div\mathbf{D} = 4\pi q \cos(\omega_e t)\, \delta(x)\delta(y)\delta(z-d), \\ D_i(x,y,z,t) = \int_{-\infty}^{0} \varepsilon_{ij}(t-\tau) E_j(x,y,z,\tau)d\tau. \end{cases} \quad (S.1)$$

The boundary conditions for Eq.(S.1) are:

$$\begin{cases} \left(D_z^{(1)} - D_z^{(2)}\right)\Big|_{z=0} = 0, \\ \left(E_x^{(1)} - E_x^{(2)}\right)\Big|_{z=0} = 0. \end{cases} \quad (S.2)$$

Here the tensor of the dielectric permittivity has the following form:

$$\varepsilon_{ij}(\omega) = \begin{bmatrix} \varepsilon_\perp & 0 & 0 \\ 0 & \varepsilon_\perp & 0 \\ 0 & 0 & \varepsilon_\parallel(\omega) \end{bmatrix}. \quad (S.3)$$

The Fourier transform on time leads to expressions:

$$D_{x,y}(x,y,z,\omega) = \varepsilon_\perp E_{x,y}(x,y,z,\omega), \quad (S.4a)$$

$$D_z(x,y,z,\omega) = \varepsilon_\parallel(\omega) E_z(x,y,z,\omega). \quad (S.4b)$$

Next, the following equation can be derived from Eq.(S.1):

$$\varepsilon_\perp \left[\frac{\partial}{\partial x}E_1(\omega) + \frac{\partial}{\partial y}E_2(\omega)\right] + \varepsilon_\parallel(\omega)\frac{\partial}{\partial z}E_3(\omega) = Re\left[\int_{-\infty}^{+\infty} 4\pi q\, exp(i\,\omega_e t)\, \delta(x)\delta(y)\delta(z-d) e^{-i\omega t} d\omega\right]. \quad (S.5)$$

Or (omitting the *Re* for the time being)

$$\varepsilon_\perp \left[\frac{\partial}{\partial x}E_1(\omega) + \frac{\partial}{\partial y}E_2(\omega)\right] + \varepsilon_\parallel(\omega)\frac{\partial}{\partial z}E_3(\omega) = 4\pi^2 q\, \delta(x)\delta(y)\delta(z-d)\delta(\omega-\omega_e). \quad (S.6)$$

And

$$\mathbf{E} = -\vec{\nabla}\varphi. \quad (S.7)$$

If we neglect the magnetic fields, since $rot\mathbf{E} = 0$, we get the following equations and boundary conditions for the scalar electrostatic potential:

$$\begin{cases} \varepsilon_e \Delta \varphi^{(1)}(\omega, x, y, z) = -4\pi^2 q \, \delta(x)\delta(y)\delta(z-d)\delta(\omega - \omega_e), & z > 0, \\ \varepsilon_\perp \Delta_\perp \varphi^{(2)}(\omega, x, y, z) + \varepsilon_\|(\omega) \frac{\partial^2}{\partial z^2} \varphi^{(2)}(\omega, x, y, z) = 0, & z < 0, \\ \varphi^{(1)}\big|_{z=0} = \varphi^{(2)}\big|_{z=0}, \\ \varepsilon_e \frac{\partial \varphi^{(1)}}{\partial z}\bigg|_{z=0} = \varepsilon_\|(\omega) \frac{\partial \varphi^{(2)}}{\partial z}\bigg|_{z=0}. \end{cases} \quad (S.8)$$

Let us do a Fourier transform on $y$ and $x$:

$$\tilde{\varphi}_n(\omega, k_1, k_2, z) = \iint_{-\infty}^{+\infty} e^{ik_1 x} e^{ik_2 y} \varphi^{(n)}(\omega, x, y, z) dx dy. \quad (S.9)$$

Here $n$ denotes the media type (1 – the ambient, 2 – the dielectric). Using this expression, from Eq.(S.8) we get the following boundary conditions:

$$\tilde{\varphi}_1(\omega, k_1, k_2, z)|_{z=0} = \tilde{\varphi}_2(\omega, k_1, k_2, z)|_{z=0}, \quad (S.10a)$$

$$\varepsilon_e \frac{\partial \tilde{\varphi}_1(\omega, k_1, k_2, z)}{\partial z}\bigg|_{z=0} = \varepsilon_\|(\omega) \frac{\partial \tilde{\varphi}_2(\omega, k_1, k_2, z)}{\partial z}\bigg|_{z=0}. \quad (S.10b)$$

The equations (S.8) can be transformed as follows:

$$\frac{\partial^2}{\partial z^2} \tilde{\varphi}_1(\omega, k_1, k_2, z) - (k_1^2 + k_2^2)\tilde{\varphi}_1(\omega, k_1, k_2, z) = $$
$$-4\pi^2 \frac{q}{\varepsilon_e} \left[\iint_{-\infty}^{+\infty} e^{ik_1 x} e^{ik_2 y} \delta(x)\delta(y) dx dy\right] \delta(z-d)\delta(\omega - \omega_e). \quad (S.11a)$$

Or in a simplified form:

$$\frac{\partial^2}{\partial z^2} \tilde{\varphi}_1(\omega, k_1, k_2, z) - (k_1^2 + k_2^2)\tilde{\varphi}_1(\omega, k_1, k_2, z) = -4\pi^2 \frac{q}{\varepsilon_e} \delta(z-d)\delta(\omega - \omega_e). \quad (S.11b)$$

The general and partial solutions of Eqs.(S.11) are:

$$\tilde{\varphi}_1(\omega, k_1, k_2, z) = C_1 e^{-kz} + \tilde{\varphi}_1^{partial}(\omega, k_1, k_2, z), \quad (S.12a)$$

$$\tilde{\varphi}_1^{partial}(\omega, k_1, k_2, z) = A \, e^{-k|z-d|}. \quad (S.12b)$$

Here $k = \sqrt{k_1^2 + k_2^2}$. The constant multiplier $A$ can be determined after the substitution of the partial solution (S.12b) in the Eq.(S.11b):

$$A k \frac{d}{dz}\left(-sign(z-d) \, e^{-k|z-d|}\right) - k^2 A \, e^{-k|z-d|} = -\frac{4\pi^2 q}{\varepsilon_e} \delta(z-d)\delta(\omega - \omega_e). \quad (S.13a)$$

If we take the derivative second time, we obtain the following equation on the constant $A$:

$$A k \left(-2 \, \delta(z-d) \, e^{-k|z-d|}\right) = -\frac{4\pi^2 q}{\varepsilon_e} \delta(z-d)\delta(\omega - \omega_e). \quad (S.13b)$$

Here we take into account, that $[sign(z-d)]^2 \equiv 1$. Due to this property, the second term cancels out the first term in the derivative. Also, it is important to note that the Dirac delta-function has the following property:

$$\delta(y-d)f(y) = \frac{f(d+0) + f(d-0)}{2} \delta(y-d). \quad (S.14)$$

Taking into account Eqs.(S.14) and (S.13b), one can obtain the formula:

$$A = \frac{4\pi^2 q}{\varepsilon_e k} \delta(\omega - \omega_e). \qquad (S.15)$$

So, we can write as follows:

$$\tilde{\varphi}_1(\omega, k_1, k_2, z) = C_1 e^{-kz} + \frac{4\pi^2 q}{\varepsilon_e k} \delta(\omega - \omega_e) e^{-k|z-d|}. \qquad (S.16)$$

The equation for the potential in the anisotropic media can be written as follows:

$$-\varepsilon_\perp (k_1^2 + k_2^2)\tilde{\varphi}_2(\omega, k_1, k_2, z) + \varepsilon_\parallel(\omega) \frac{\partial^2}{\partial z^2} \tilde{\varphi}_2(\omega, k_1, k_2, z) = 0. \qquad (S.17)$$

The general solution:

$$\tilde{\varphi}_2(\omega, k_1, k_2, z) = C_2 e^{kz\,\zeta(\omega)}, \qquad (S.18a)$$

Where the integral convergency conditions is

$$Re\,[\zeta(\omega)] > 0. \qquad (S.18b)$$

The dielectric anisotropy factor $\zeta(\omega) = \sqrt{\frac{\varepsilon_\perp}{\varepsilon_\parallel(\omega)}}$. The constants $C_{1,2}$ can be determined from the boundary conditions:

$$C_1 + \frac{4\pi^2 q}{\varepsilon_e k} \delta(\omega - \omega_e) e^{-kd} = C_2, \qquad (S.19a)$$

$$\varepsilon_e \left[ -C_1 k + \frac{4\pi^2 q}{\varepsilon_e} \delta(\omega - \omega_e) e^{-kd} \right] = C_2 \varepsilon_\parallel(\omega)\, k\zeta(\omega). \qquad (S.19b)$$

So, the constants have the following values:

$$C_1 = -\frac{4\pi^2 q}{\varepsilon_e k} \delta(\omega - \omega_e) e^{-d\,k} \frac{\varepsilon_\parallel(\omega)\, \zeta(\omega) - \varepsilon_e}{\varepsilon_e + \varepsilon_\parallel(\omega)\, \zeta(\omega)}, \qquad (S.20a)$$

$$C_2 = \frac{4\pi^2 q}{\varepsilon_e\, k} \delta(\omega - \omega_e) e^{-d\,k} \frac{2\,\varepsilon_e}{\varepsilon_e + \varepsilon_\parallel(\omega)\, \zeta(\omega)}. \qquad (S.20b)$$

Now we can write the Fourier images of the potentials as follows:

$$\tilde{\varphi}_1(\omega, k_1, k_2, z) = \frac{4\pi^2 q}{\varepsilon_e k} \left[ e^{-k|z-d|} - e^{-kd} f(\omega) e^{-kz} \right] \delta(\omega - \omega_e), \qquad (S.21a)$$

$$\tilde{\varphi}_2(\omega, k_1, k_2, z) = \frac{4\pi^2 q}{\varepsilon_e k}\, e^{-kd} g(\omega) e^{kz\,\zeta(\omega)} \delta(\omega - \omega_e). \qquad (S.21b)$$

Let us start doing the inverse Fourier transform. At first, we can do the easiest one, on the frequency. Let us introduce the following functions:

$$f(\omega) = \frac{\sqrt{\varepsilon_\perp \varepsilon_\parallel(\omega)} - \varepsilon_e}{\varepsilon_e + \sqrt{\varepsilon_\perp \varepsilon_\parallel(\omega)}}, \qquad (S.22a)$$

$$g(\omega) = \frac{2\,\varepsilon_e}{\varepsilon_e + \sqrt{\varepsilon_\perp \varepsilon_\parallel(\omega)}}. \qquad (S.22b)$$

Using Eqs.(S.22) we derive the following expressions:

$$\tilde{\psi}_1(t, k_1, k_2, z) = \frac{2\pi q}{\varepsilon_e k} \left[-e^{-kd} f(\omega_e) e^{-kz} + e^{-k|z-d|}\right] e^{i\omega_e t}, \quad \text{(S.23a)}$$

$$\tilde{\psi}_2(t, k_1, k_2, z) = \frac{2\pi q}{\varepsilon_e k} e^{-kd} g(\omega_e) e^{i\omega_e t} e^{kz\,\zeta(\omega)}. \quad \text{(S.23b)}$$

Next, we can inverse the transformation on the coordinates. Using the integral $\int_0^\infty \frac{\cos(k\beth)}{\sqrt{k^2+\vartheta^2}} e^{-\aleph\sqrt{k^2+\vartheta^2}} dk = K_0(\vartheta\sqrt{\beth^2+\aleph^2})$. We need to transfer to a polar coordinate system (for the wave-vectors):

$$\varphi^{(n)}(t,x,y,z) = \frac{1}{4\pi^2} \iint_{-\infty}^{+\infty} e^{-ik_1 x} e^{-ik_2 y} \tilde{\psi}_n(t,k,z) dk_1 dk_2 =$$
$$\frac{1}{4\pi^2} \int_0^\infty \tilde{\psi}_n(t,k,z) k\, dk \int_0^{2\pi} e^{-ik\cos(\phi)x} e^{-ik\sin(\phi)y} d\phi. \quad \text{(S.24)}$$

The angular part:

$$\int_0^{2\pi} e^{-ik\cos(\phi)x} e^{-ik\sin(\phi)y} d\phi = \int_0^{2\pi} e^{[-ikr\cos(\phi - \arccos(\phi_0))]} d\phi = 2\pi J_0(kr). \quad \text{(S.25)}$$

Here the polar radius $r = \sqrt{x^2+y^2}$ is introduced. So, we can write the following expressions:

$$\varphi^{(n)}(t,r,z) = \frac{1}{2\pi} \int_0^\infty \tilde{\psi}_n(t,k,z) J_0(k\,r) k\, dk. \quad \text{(S.26)}$$

The final formulae:

$$\varphi^{(1)}(t,r,z) = \frac{q}{\varepsilon_e} \int_0^\infty e^{i\omega_e t} \left[-e^{-kd} f(\omega_e) e^{-kz} + e^{-k|z-d|}\right] J_0(k\,r)\, dk, \quad \text{(S.27a)}$$

$$\varphi^{(2)}(t,r,z) = \frac{q}{\varepsilon_{ext}} \int_0^\infty e^{-kd} g(\omega_e) e^{i\omega_e t} e^{kz\,\zeta(\omega)} J_0(k\,r)\, dk. \quad \text{(S.27b)}$$

Using the integral:

$$\int_0^\infty e^{-kz} J_0(ka) dk = \frac{1}{\sqrt{z^2+a^2}}. \quad \text{(S.28)}$$

we can write the potential as:

$$\varphi^{(1)}(t,r,z) = \frac{q}{\varepsilon_e} Re\left[\frac{e^{i\omega_e t}}{\sqrt{(z-d)^2+r^2}} - f(\omega_e) \frac{e^{i\omega_e t}}{\sqrt{(z+d)^2+r^2}}\right], \quad \text{(S.29a)}$$

$$\varphi^{(2)}(t,r,z) = \frac{q}{\varepsilon_e} Re\left[\frac{g(\omega_e) e^{i\omega_e t}}{\sqrt{r^2 + \left(d - z\sqrt{\frac{\varepsilon_\perp}{\varepsilon_\parallel(\omega_e)}}\right)^2}}\right], \quad \text{(S.29b)}$$

In what follows we consider the expressions for the dielectric permittivity allowing for the frequency dispersion in the medium 2:

$$\varepsilon_e = constant, \quad \varepsilon_\perp = constant, \quad \varepsilon_\parallel(\omega) = \chi_\infty \frac{\omega_{lo}^2 - \omega^2 - 2i\gamma\omega}{\omega_{to}^2 - \omega^2 - 2i\gamma\omega}. \quad \text{(S.30)}$$

The z-component of the $\varepsilon_\parallel(\omega)$ can be separated into an imaginary and real parts:

$$\varepsilon_\parallel(\omega) = \varepsilon' + i\,\varepsilon'', \quad \text{(S.31a)}$$

$$\varepsilon' = \chi_\infty \frac{(\omega_{lo}^2-\omega^2)(\omega_{to}^2-\omega^2) + 4\gamma^2\omega^2}{(\omega_{to}^2-\omega^2)^2 + 4\gamma^2\omega^2}, \quad \varepsilon'' = \chi_\infty \frac{2\gamma\omega(\omega_{lo}^2-\omega_{to}^2)}{(\omega_{to}^2-\omega^2)^2 + 4\gamma^2\omega^2}. \quad \text{(S.31b)}$$

The complex effective permittivity is:

$$\sqrt{\varepsilon_\perp \varepsilon_\|(\omega)} = \sqrt{\varepsilon_\perp \sqrt{(\varepsilon'^2 + \varepsilon''^2)}} \exp\left[i\left(\frac{1}{2}\arctan\left(\frac{\varepsilon''}{\varepsilon'}\right) + \pi n\right)\right], \quad n = 1,2. \tag{S.32}$$

Let us substitute this expression into Eqs.(S.29a) and (S.29b):

$$f(\omega) = \frac{\sqrt{\varepsilon_\perp \varepsilon_\|(\omega_e)} - \varepsilon_e}{\varepsilon_e + \sqrt{\varepsilon_\perp \varepsilon_\|(\omega_e)}} = \frac{\sqrt{\varepsilon_\perp \sqrt{(\varepsilon'^2 + \varepsilon''^2)}} \exp\left[i\left(\frac{1}{2}\arctan\left(\frac{\varepsilon''}{\varepsilon'}\right) + \pi n\right)\right] - \varepsilon_e}{\varepsilon_e + \sqrt{\varepsilon_\perp \sqrt{(\varepsilon'^2 + \varepsilon''^2)}} \exp\left[i\left(\frac{1}{2}\arctan\left(\frac{\varepsilon''}{\varepsilon'}\right) + \pi n\right)\right]}. \tag{S.33}$$

Similarly the function

$$g(\omega) = \frac{2\varepsilon_e}{\varepsilon_e + \sqrt{\varepsilon_\perp \varepsilon_\|(\omega)}} = \frac{2\varepsilon_e}{\varepsilon_e + \sqrt{\varepsilon_\perp \sqrt{(\varepsilon'^2 + \varepsilon''^2)}} \exp\left[i\left(\frac{1}{2}\arctan\left(\frac{\varepsilon''}{\varepsilon'}\right) + \pi n\right)\right]}. \tag{S.34}$$

Next, let us introduce the following functions:

$$\sigma(\omega_e) = \sqrt{\varepsilon_\perp \sqrt{(\varepsilon'^2 + \varepsilon''^2)}}, \tag{S.35a}$$

$$\tau_n(\omega_e) = \left(\frac{1}{2}\arctan\left(\frac{\varepsilon''}{\varepsilon'}\right) + \pi n\right). \tag{S.35b}$$

We need to choose the leaf (given by $n$) from the condition (S.18b):

$$Re\sqrt{\frac{\varepsilon_\perp}{\varepsilon_\|(\omega)}} = Re\sqrt{\frac{\varepsilon_\perp}{\varepsilon' + i\varepsilon''}} = \sqrt{\frac{\varepsilon_\perp}{\varepsilon'^2 + \varepsilon''^2}} Re\left[\sqrt{\varepsilon' - i\varepsilon''}\right] > 0, \tag{S.36a}$$

$$Re\left[\sqrt{\varepsilon' - i\varepsilon''}\right] = Re\left[\sqrt{\varepsilon_\perp \sqrt{(\varepsilon'^2 + \varepsilon''^2)}} \exp\left[i\left(-\frac{1}{2}\arctan\left(\frac{\varepsilon''}{\varepsilon'}\right) + \pi n\right)\right]\right] > 0. \tag{S.36b}$$

After some simplifications,

$$(-1)^n \cos\left(\frac{1}{2}\arctan\left(\frac{\varepsilon''}{\varepsilon'}\right)\right) > 0. \tag{S.36c}$$

So, the sign of the $\cos\left(\frac{1}{2}\arctan\left(\frac{\varepsilon''}{\varepsilon'}\right)\right)$ defines the number of the leaf we use. Hence, to solve this problem, we use that $\cos(\tau_n(\omega_e)) \to |\cos(\tau(\omega_e))|$. Here

$$\tau(\omega_e) = \frac{1}{2}\arctan\left(\frac{\varepsilon''}{\varepsilon'}\right). \tag{S.37}$$

Using simple trigonometry, we write as follows:

$$\cos(\omega_e t \pm \tau_n(\omega_e)) = \cos(\omega_e t)|\cos(\tau(\omega_e))| \mp \sin(\omega_e t)\sin(\tau(\omega_e)). \tag{S.38}$$

Using these formulae, the following contour maps of the potential were obtained:

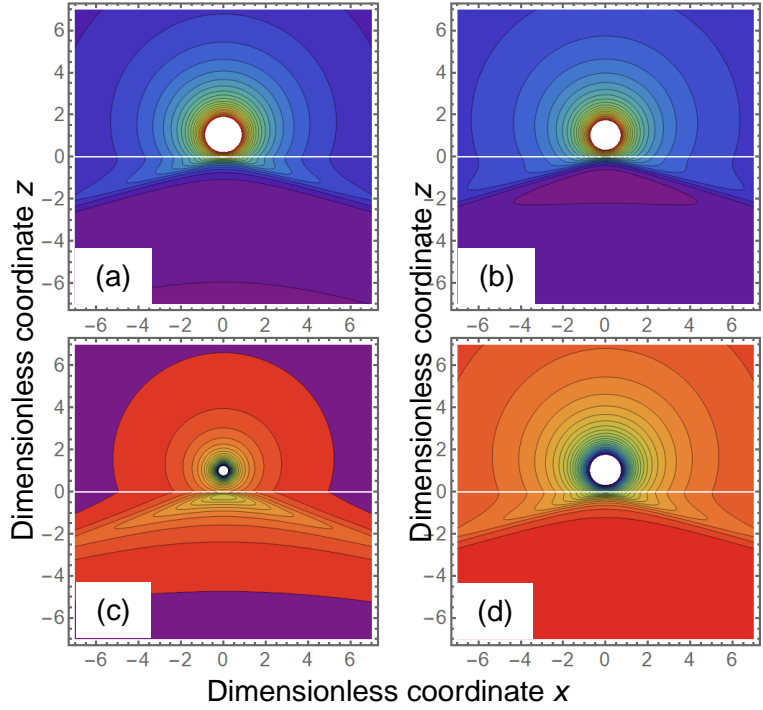

**Figure S1**. The contour maps of the potential in different moments. The time moments are 0.01 **(a)**, 0.1 **(b)**, 0.3 **(c)**, 0.5 **(d)** of the oscillation period. The coordinates are divided by the effective distance $d$.

Let us evaluate the dielectric loss in the media. The general expression for the power of the losses is [i]:

$$\frac{d\mathcal{E}}{dt} = -\frac{1}{4\pi} \iiint_V \vec{E} \frac{\partial \vec{D}}{\partial t} dV. \qquad (S.39)$$

The full energy per unit length can be written as follows:

$$E = \Delta x \int_{-\infty}^{+\infty} dt \iint_{-\infty}^{+\infty} \rho_Q dy dz. \qquad (S.40a)$$

Let us consider the general form of the spatial density in (S.40a), implying that we use a Fourier integral view of the electric displacement and the field:

$$\rho_E = \int_{-\infty}^{\infty} dt \iint_{-\infty}^{+\infty} \rho_Q(\omega_1, \omega_2, t) d\omega_1 d\omega_2 = -\frac{1}{4\pi} \int_{-\infty}^{+\infty} dt \iint_{-\infty}^{+\infty} \frac{1}{2} \Big[ i\omega_2 \vec{E}_\omega(\omega_1) \vec{D}_\omega^*(\omega_2) e^{-i(\omega_1-\omega_2)t} - i\omega_2 \vec{E}_\omega^*(\omega_1) \vec{D}_\omega(\omega_2) e^{-i(\omega_2-\omega_1)t} \Big] d\omega_1 d\omega_2 = -\frac{1}{8\pi} \iint_{-\infty}^{+\infty} \frac{1}{2} \Big[ i\omega_2 \vec{E}_\omega(\omega_1) \vec{D}_\omega^*(\omega_2) - i\omega_2 \vec{E}_\omega^*(\omega_1) \vec{D}_\omega(\omega_2) \Big] \delta(\omega_2 - \omega_1) d\omega_1 d\omega_2 = -\frac{1}{8\pi} \int_{-\infty}^{+\infty} \Big[ i\omega \vec{E}_\omega(\omega) \vec{D}_\omega^*(\omega) - i\omega_2 \vec{E}_\omega^*(\omega) \vec{D}_\omega(\omega) \Big] d\omega. \qquad (S.40b)$$

Here the integrant

$$\rho_Q(\omega_1, \omega_2, t) = -\frac{1}{8\pi} \Big[ \vec{E}_\omega(\omega_1) e^{-i\omega_1 t} \frac{\partial}{\partial t} \Big( \vec{D}_\omega^*(\omega_2) e^{-i\omega_2 t} \Big) + c.c. \Big]. \qquad (S.40c)$$

Taking into account the view of the dielectric permittivity, one can obtain that

$$\rho_E = -\frac{1}{8\pi} \int_{-\infty}^{+\infty} \Big[ i\omega \vec{E}_\omega(\omega)(\varepsilon' - i\varepsilon'') \vec{E}_\omega^*(\omega) - i\omega \vec{E}_\omega^*(\omega)(\varepsilon' + i\varepsilon'') \vec{E}_\omega(\omega) \Big] d\omega. \qquad (S.41a)$$

Considering the symmetry properties of the dielectric permittivity one can write that

$$\rho_E = -\frac{1}{8\pi}\int_{-\infty}^{+\infty}\left[i\omega|\vec{E}_\omega(\omega)|^2(-i\,\varepsilon'') - i\omega|\vec{E}_\omega(\omega)|^2 i\,\varepsilon''\right]d\omega = -\frac{1}{2\,\pi}\int_0^{+\infty}\omega|\vec{E}_\omega(\omega)|^2\varepsilon''d\omega. \quad \text{(S.41b)}$$

We can find the field as follows:

$$\vec{E}_\omega(\omega) = -\vec{\nabla}\psi(\omega,\vec{r}). \quad \text{(S.42)}$$

So, the final formula for the density is shown below:

$$\rho_E = -\frac{1}{2\,\pi}\int_0^{+\infty}\omega|\vec{\nabla}\psi(\omega,\vec{r})|^2\varepsilon''d\omega = -\frac{1}{2\,\pi}\int_0^{+\infty}|\vec{\nabla}\psi(\omega,\vec{r})|^2\sigma(\omega)d\omega. \quad \text{(S.43)}$$

The above formula is written for purely conductive losses, $\varepsilon'' = \frac{\sigma}{\omega}$. In our case

$$\psi(\omega,\vec{r}) \sim \frac{1}{\sqrt{r^2 + \left(d - z\sqrt{\varepsilon_\perp/\varepsilon_\parallel(\omega)}\right)^2}}. \quad \text{(S.44)}$$

The $\psi$-gradient can be written as follows:

$$\vec{e}_r \frac{r}{\sqrt{\left(r^2+(d-z\,\zeta(\omega))^2\right)^3}} + \vec{e}_z \frac{-(d-z\,\zeta(\omega))\zeta(\omega)}{\sqrt{\left(r^2+(d-z\,\zeta(\omega))^2\right)^3}}. \quad \text{(S.45)}$$

The module squared of the gradient is:

$$\frac{r^2 + |d-z\,\zeta(\omega)|^2|\zeta(\omega)|^2}{\left|\left(r^2+(d-z\,\zeta(\omega))^2\right)^3\right|}. \quad \text{(S.46)}$$

---

[i] Landau, Solid state electrodynamics, page 306, formula (56,15)